\documentclass[aps,prl,twocolumn,floatfix]{revtex4}
\usepackage{graphicx,epsfig,array,amsmath,bm}

\begin{document}

\title{Theory of High T$_c$ Ferrimagnetism in a Multi-orbital Mott Insulator}

\author{O. Nganba Meetei$^1$, Onur Erten$^1$, Mohit Randeria$^1$, Nandini Trivedi$^1$ and Patrick Woodward$^2$}

\affiliation{$^1$Department of Physics, The Ohio State University, Columbus, Ohio 43210, USA\\ 
$^2$ Department of Chemistry, The Ohio State University, Columbus, Ohio 43210, USA } 
\begin{abstract}
We propose a model for
the multi-orbital material Sr$_2$CrOsO$_6$ (SCOO), an insulator with remarkable magnetic properties and
the highest $T_c \simeq 725$~K among {\em all} perovskites with a net moment.
We derive a new criterion for the Mott transition $(\widetilde{U}_{1} \widetilde{U}_{2})^{1/2}>2.5W$ using slave rotor mean field theory,
where $W$ is the bandwidth and $\widetilde{U}_{1(2)}$ are the effective Coulomb interactions on Cr(Os) 
including Hund's coupling. We show that SCOO is a Mott insulator, where the large Cr $\widetilde{U}_{1}$
compensates for the small Os $\widetilde{U}_{2}$. The spin sector is described by a 
frustrated antiferromagnetic Heisenberg model that naturally explains the net moment arising from canting 
and also the observed non-monotonic magnetization $M(T)$.
We predict characteristic magnetic structure factor peaks that can be probed by neutron experiments.
\end{abstract}

\maketitle

Transition metal oxides~\cite{Imada} with partially filled 3d and 4d shells have dominated materials research in 
past decades, leading to such spectacular phenomena as high $T_c$ superconductivity and
colossal magnetoresistance. We expect the next revolution to involve 5d oxides where the interplay 
of spin-orbit coupling (SOC) and strong correlations can lead to novel phases of matter. 
Recent experimental efforts have focused on iridium oxides ~\cite{kim_2009,liu_2011,matsuhira_2007,okamoto_2007},
driven by the possibility of exotic phases~\cite{khaliullin_2009,ashvin_2011,Balents_SR,senthil_2011}. 
Here we explore other equally promising experimental systems, the 5d oxides
containing osmium that have received far less attention and provide a rich area for
investigation. 

We focus on the double perovskite (DP) family of oxides with the formula A$_2$BB$^\prime$O$_6$. 
A is an alkali or alkaline earth metal while B and B$^\prime$ are two transition metals (TMs) arranged 
on a 3D checkerboard lattice. With two TM ions, the range of properties span metals to insulators, 
ranging from ferromagnets, antiferromagnets, ferroelectrics, multiferroics, to spin liquids~\cite{Ibarra_2007, Kobayashi, Balents_DP1, Balents_DP2}.

The most studied DP is Sr$_2$FeMoO$_6$, a half metallic ferrimagnet with 
$T_c\approx$ 420K~\cite{Kobayashi,DDSarma_SFMO}. 
It is now well-understood that a generalized double exchange mechanism can explain  
ferromagnetism with the scale of $T_c$ set by the kinetic energy of itinerant electrons~\cite{our_prl}.
From this perspective, the observation of an even higher $T_c\approx$ 725K,
the highest $T_c$ amongst {\it all} perovskites with a net moment,
 in an insulator 
Sr$_2$CrOsO$_6$ (SCOO)~\cite{Krockenberger} is rather puzzling. 

There have been several important density functional theory (DFT) calculations of SCOO
\cite{Krockenberger,Pickett1,Kubler,DDSarma}, nevertheless many puzzling questions
remain open.
(i) Why is SCOO an insulator~\cite{note1}? It is not a band insulator, since the bands are partially filled.
It is not obviously a Slater insulator, given the large
moment observed on Cr.  It is not {\it a priori} clear how it can be a Mott insulator either, given 
the weak correlations on Os relative to the large bandwidth of 5d orbitals. 
(ii) Why is there a net moment, given that both Cr and Os are in $d^3$ configurations?  
What is the role of SOC on Os? 
DFT calculations~\cite{DDSarma,Pickett1} find a net moment only if SOC is included, but
XMCD experiments~\cite{Krockenberger} show insignificant Os orbital moment.
(iii) What sets the scale for the high magnetic $T_c \simeq 725$K?  DFT does not give a 
clear answer~\cite{note2}. 
(iv)  Why is the magnetization $M(T)$ a non-monotonic function ~\cite{Krockenberger} of $T$?
Finite temperature properties are, of course, very hard for DFT to address.

In the Letter we gain insight into all these questions within a theoretical framework 
that describes the hierarchy of energy scales in the charge and spin sectors. 
\\
(1) We use a slave-rotor mean field theory to analyze the charge sector of SCOO and show that it
is a multi-orbital Mott insulator. We derive a new criterion for the Mott transition for this system: 
$(\widetilde{U}_{1} \widetilde{U}_{2})^{1/2}>2.5W$ where W is the bandwidth and $\widetilde{U}_{1(2)}$ the effective Coulomb
repulsion on Cr(Os), which includes the important effects of Hund's coupling in addition to Hubbard $U$. 
Thus a small $\widetilde{U}_2$ on Os is compensated by a strong $\widetilde{U}_1$ on Cr and drives the system into a Mott insulating state.
\\
(2) We find that the orbital angular momentum on Os is quenched, and the Os moment is purely of spin origin, in agreement with XMCD 
measurements~\cite{Krockenberger}. 
\\
(3) We show that the effective spin Hamiltonian for SCOO is a frustrated Heisenberg model with antiferromagnetic
exchange between nearest neighbor Cr-Os as well as next-nearest neighbor Os-Os spins. We find that the net magnetic moment
$M(0)$ arises, not because of SOC, but rather because of a canted magnetic ground state, found using both Monte Carlo (MC)
simulations and a variational approach.
\\
(4) The scale for the high $T_c$ is set by a $J_1 = 35$ meV superexchange between Cr-Os near neighbors.
\\
(5) We find that we can obtain an unusual non-monotonic $M(T)$, in agreement with magnetization and neutron 
experiments~\cite{Krockenberger}, due to a subtle frustration effect between Cr-Os $J_1$ and Os-Os $J_2$. 
\\
(6) Finally, we predict distinct magnetic structure factor peaks for our canted ground state.
These can be measured by neutrons and provide a ``smoking gun" test of our theory.

\smallskip

\noindent {\em Hamiltonian:}
The cubic crystal field splits $d$-orbitals and only the t$_{2g}$ orbitals are occupied, with both Cr and Os in $d^3$ configurations.
To investigate the metal-insulator transition, we consider
\begin{eqnarray}
\nonumber
H&=&-t \sum_{\langle ij \rangle, \alpha} (d_{i\alpha}^\dagger c_{j\alpha}+h.c)-t^\prime \sum_{\langle \langle ij  \rangle \rangle, \alpha} (c_{i\alpha}^\dagger c_{j\alpha}+h.c)\\ \nonumber
&&+\frac{\widetilde{U}_{1}}{2}\sum_i \Big[ \sum_{\alpha} n_{1i\alpha} \Big]^2+ \frac{\widetilde{U}_{2}}{2}\sum_i \Big[ \sum_{\alpha} n_{2i\alpha} \Big]^2 \\
&&-\mu \sum_{i,\alpha}(n_{1i\alpha}+n_{2i\alpha})
-\Delta \sum_{i,\alpha}(n_{1i\alpha}-n_{2i\alpha}). \label{full_H}
\end{eqnarray}
This has orbitally-symmetric onsite Coulomb interactions that make the problem analytically tractable and provide insight,
while capturing the essence of strong correlations in a multi-band system. 
Hund's coupling will be discussed below. 
Here $d_{i\alpha}^\dagger$ ($d_{i\alpha}$) and $c_{i\alpha}^\dagger$ ($c_{i\alpha}$) are the creation (annihilation) operators 
on the $i^{th}$ Cr and Os sites respectively. $n_{1i\alpha}$ ($n_{2i\alpha}$) are the number operators on the $i^{th}$ Cr (Os) site. 
Note that the label $\alpha$ includes both spin and t$_{2g}$ orbital index. 
$t$ and $t^{\prime}$ are the hopping amplitude between neighboring Cr-Os and Os-Os sites.  
 The {\it effective} on-site Coulomb interaction $\widetilde{U}_{1(2)}$ is related to the atomic charge gap on Cr(Os),
as discussed in detail below.  The chemical potential is $\mu$ and $\Delta$, the charge transfer gap between Cr and Os.  

\begin{figure}[!t]
\vspace{.2cm}
\centerline{
\includegraphics[width=8.5cm]{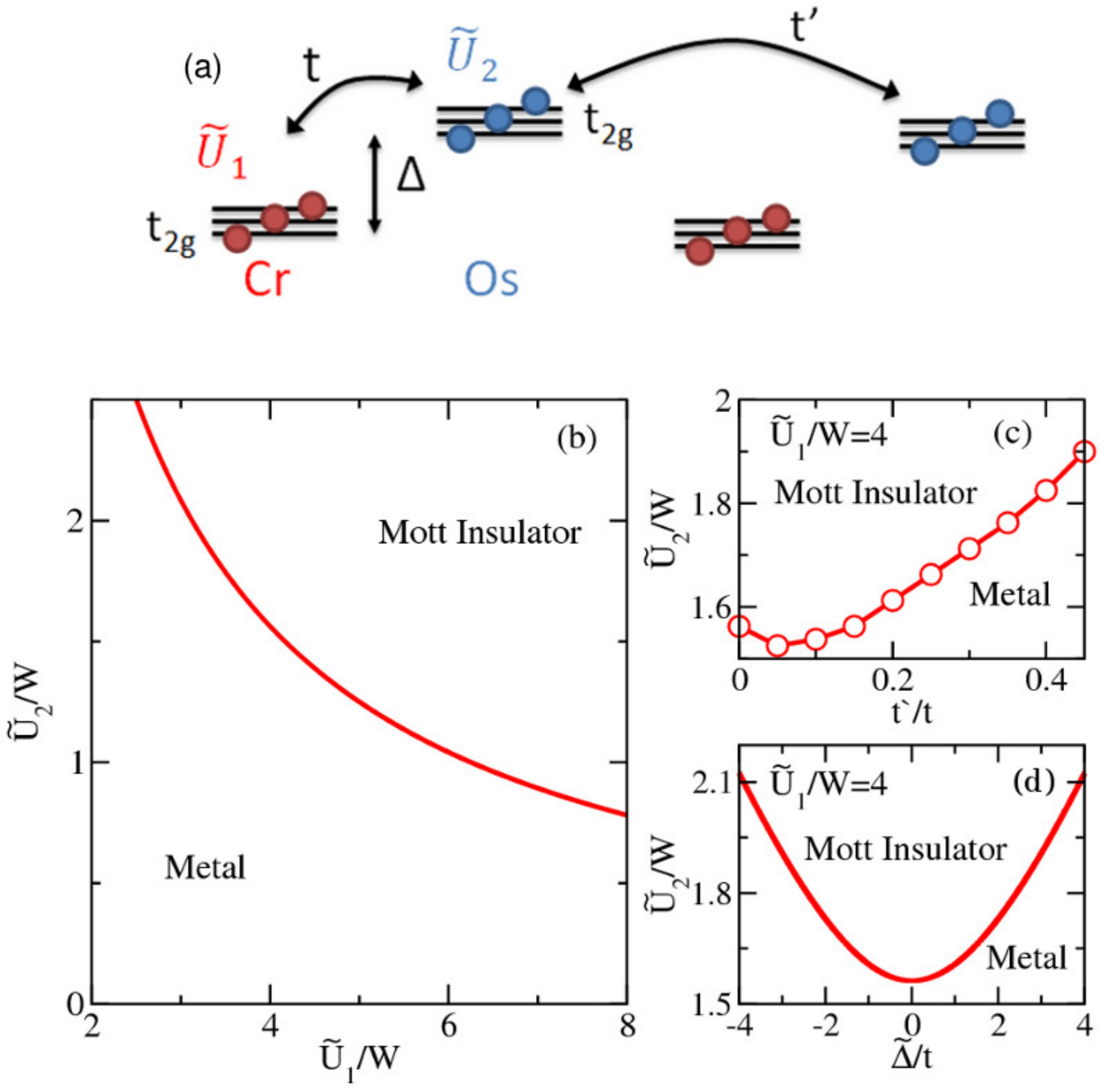}
}
\vspace{.2cm}
\caption{ Phase diagram obtained from slave-rotor mean field theory.  
(a) Schematic of Cr and Os orbitals and relevant energies. 
(b) Phase boundary of eq.~(\ref{eq:Mott_criterion}) for the particle-hole symmetric case 
$t^\prime=0$, $\Delta=\Delta_0$, and $\mu=\mu_0$; see text. Combination of small $\widetilde{U}_2$ on Os and large $\widetilde{U}_1$  on Cr can drive the system Mott insulating. 
Effect of (c) Os-Os hopping $t^\prime$ and (d) Cr-Os charge-transfer energy $\Delta$.
}
\label{f1}
\end{figure}

\noindent {\em {Multi-band Mott criterion:}} 
We investigate the charge sector of  (\ref{full_H}) using the slave-rotor mean field (MF) theory~\cite{Florens}.
This method captures the qualitative physics of dynamical mean field theory, and
is well suited to studying the multi-band Mott transition; it has recently been used for various other 
multi-band system including pyroclores and iron pnictides~\cite{Balents_SR, Lee_SR, Fiete_SR}. 
We decompose the physical electron into an O(2) rotor that carries the charge and a spinon that carries the 
spin and orbital degrees of freedom. We write $d_{i\alpha}^{\dagger}=f_{i\alpha}^\dagger e^{i\theta_i}$ 
and $c_{i\alpha}^{\dagger}=g_{i\alpha}^\dagger e^{i\phi_i}$ where $f_i^\dagger$($f_i$) and $g_i^{\dagger}$($g_i$) are the 
fermonic spinon creation(annihilation) operators on the $i^{th}$ Cr and Os sites respectively. 
The bosonic rotor operator $\theta_i$($\phi_i$)  is conjugate to the charge on Cr(Os) at site $i$. 
Substituting into (\ref{full_H}) and performing a MF decomposition of the hopping terms, 
we obtain spinon and rotor Hamiltonians that we solve self-consistently.

We calculate $\langle \cos\theta\rangle$ and $\langle \cos\phi\rangle$ which are used as diagnostics of
the metal-insulator transition. When both quantities vanish, the system is a Mott insulator with suppressed charge fluctuations,
while non-zero values lead to a metal. For simplicity, we first work at the particle-hole (p-h) symmetric point:
$\mu\equiv\mu_0=3(\tilde{U}_1+\tilde{U}_2)/2$, $\Delta\equiv\Delta_0 = 3(\tilde{U}_1-\tilde{U}_2)/2$ and $t^{\prime}=0$, 
where we can obtain analytical results~\cite{supp}. We find 
\begin{equation}
(\widetilde{U}_{1} \widetilde{U}_{2})^{1/2}>2.5W, \label{eq:Mott_criterion}
\end{equation}
where $W=8t$ is the t$_{2g}$ bandwidth. This new Mott criterion generalizes the well-known result $U\gtrsim W$.

Deviations from the p-h symmetric point, relevant for real materials, require numerical
solution of the slave rotor mean field equations. In general,
the right-hand side of (\ref{eq:Mott_criterion}) is replaced by the 
the average kinetic energy, which depends on the filling and band structure~\cite{supp}.

The resulting phase diagram is shown in Fig.~\ref{f1}. 
We see from Fig.~\ref{f1}(c) that increasing $t^{\prime}$ 
favors the metallicity by introducing an additional route for gaining kinetic energy.  
Fig.~\ref{f1}(d) shows that increasing the deviation from the symmetric point $\tilde{\Delta}\equiv \Delta-\Delta_0$,  
for a fixed $\mu=\mu_0$ and $t^{\prime}=0$, also favors metallicity as it increases charge fluctuations 
within a unit cell\cite{supp}.

\smallskip

\noindent {\em{Determining the effective U:}}
In order to apply our results to SCOO, we must relate the effective interaction $\tilde{U}_{1(2)}$ to 
material parameters: the Hubbard $U$, Hund's coupling $J_H$ and SOC $\xi$. 
This is done by matching the excitation energy in the atomic limit of (\ref{full_H}) 
to that of the more general atomic Hamiltonian 
\begin{eqnarray}
H^{at}_{a}&=&(U_a-3J_{Ha})\frac{\hat{N_a}(\hat{N_a}-1)}{2}-2J_{Ha}\left(\mathbf{S}_a\right)^2  \nonumber \\
          && +\frac{1}{2} J_{Ha} \left(\mathbf{T}_a\right)^2-\lambda_a \mathbf{T}_a\cdot\mathbf{S}_a -\mu_a\hat{N}_a. \label{eq:atomicH}
\end{eqnarray}   
Here the label $a=1,2$ indicates Cr and Os sites respectively. $\hat{N}$ is the total number operator, 
$\mathbf{S}$ the total spin operator, and 
$\mathbf{T}$ the total orbital angular momentum operator in the t$_{2g}$ manifold,  
The effective SOC $\lambda$ in a given $TS$ manifold is a function of $T$, $S$ and $\xi$ \cite{supp}. 
We work with a $d^3$ atomic ground state. 

For the Cr site, we ignore SOC and find the atomic excitation gap
to obtain $\widetilde{U}_1=U_1+2J_{H1}$~\cite{supp}. The effective repulsion
is enhanced by Hund's coupling, well known~\cite{Cox_1998,Georges_PRL} for half-filled orbitals.
On the Os site, we include $\lambda_2$ and find $\widetilde{U}_2=U_2+2J_{H2}-3\xi_2/2$ \cite{supp},
so that SOC reduces the atomic charge gap.

We now can use our Mott criterion for SCOO.
For Cr, we use $U_1\approx 5$ eV and $J_{H1}\approx 1$ eV \cite{Imada}.
For Os, we use $U_2\approx 2$ eV, $J_{H2}\approx 0.35$ eV and $\xi_2\approx 0.3$ eV \cite{pickett_2007}.
Thus $(\widetilde{U}_{1} \widetilde{U}_{2})^{1/2} \approx 4$ eV which is larger than the critical value 
$2.5 W \approx 3.75$ eV using the band theory estimate $W\approx 1.5$ eV \citep{Pickett1}.
We thus believe that SCOO is a Mott insulator.
We emphasize that (\ref{eq:Mott_criterion}) is a conservative criterion for a Mott insulator
for two reasons. First, MF theory overestimates the stability of the metallic phase 
(which has long range order in the rotor sector). Second, the slave rotor approach neglects
magnetic order, which would further help charge localization.

It is worthwhile to check eq.~(\ref{eq:Mott_criterion}) for a
ternary perovskite NaOsO$_3$, where it reduces to $\tilde{U} > 2.5 W$. 
NaOsO$_3$ does not satisfy this inequality and is thus on the metallic side of the transition.
This is in agreement with experiments~\cite{Shi_PRB2009,Calder_arxiv2012} that show metallic behavior 
in the paramagnetic state, with spin density wave magnetism at low $T$. 
On the other hand, for a DP like SCOO, 
{\textit{it is the geometric mean of $\widetilde{U}_1$ and $\widetilde{U}_2$ that determines the effective strength of correlation}}. 
Thus the small onsite $\widetilde{U}_2$ on the $5d$ site Os is compensated by the large $\widetilde{U}_1$ on the $3d$ 
site like Cr to drive the system Mott insulating. 

\begin{figure}[!t]
\centering
\includegraphics[width=8.5cm]{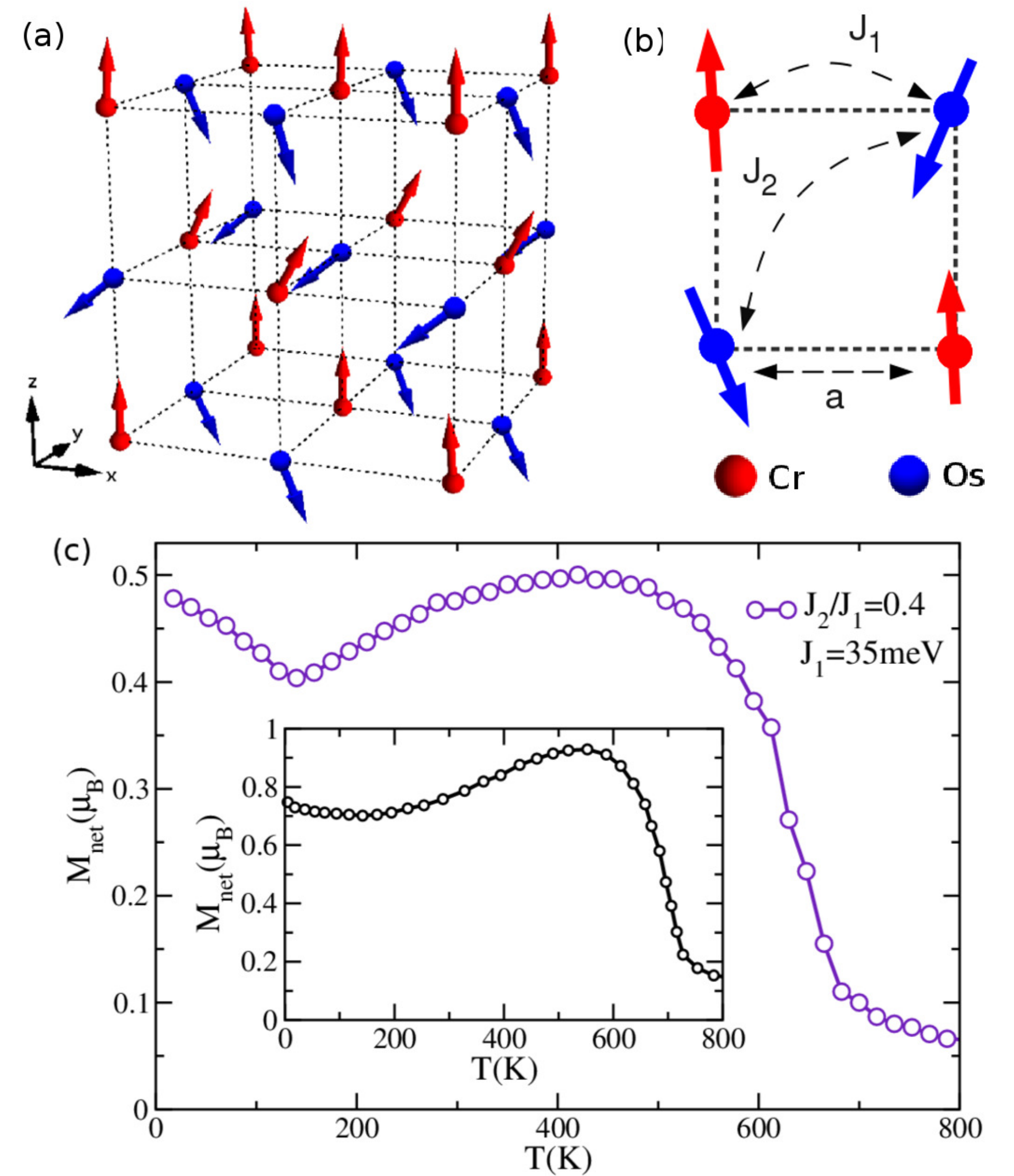}
\caption{(a) Canted ground state obtained from MC simulations ($4096$ spins, $J_2/J_1=0.4$).
(b) Schematic showing the antiferromagnetic interactions.
(c) $M(T)$ for $J_2/J_1=0.4$. By comparing with experimental $T_c$ we obtain $J_1$=35 meV.
Note that $M(T\!=\!0)$ is finite and $M(T)$ has an unusual non-monotonic behavior in agreement with experimental data
in inset~\cite{Krockenberger}. }
\label{f3}
\end{figure}

\smallskip

\noindent {\em{Effective Spin Hamiltonian:}}
Having shown that charge is localized, we turn to the spin sector.
For simplicity, we take a strong coupling approach, although in reality SCOO is likely
an intermediate coupling material (see Concluding section).
Given the large $J_H$ on Cr, the three spins align to yield S=3/2. 
We might expect that a large SOC $\lambda \bf{l} \cdot \bf{S}$ on Os 
might split the 5d $t_{2g}$ orbitals into $J=3/2$ and $J=1/2$ manifolds. However, in the special case of a $d^3$ configuration, 
the orbital angular momentum is quenched. Thus we also get a
S=3/2 spin on Os. While the absence of an orbital moment on Os is consistent with XMCD data~\cite{Krockenberger}, we must
now explain the net moment, since SOC cannot be responsible for this.

The effective spin Hamiltonian is given by 
\begin{eqnarray}
 H_{\rm eff}= J_1 \sum_{\langle ij \rangle} {\mathbf{S}_i^{1} \cdot \mathbf{S}_j^{2}} + J_2 \sum_{\langle\langle ij \rangle\rangle} {\mathbf{S}_i^{2} \cdot \mathbf{S}_j^{2}}.
\label{eq:spinH}
\end{eqnarray}
Here $\mathbf{S}_i^{1(2)}$ is a classical S=3/2 spin on the $i^{\rm th}$ Cr(Os) site;
see Fig \ref{f3}(b). 
Given the general rules of superexchange for half-filled orbitals, 
$J_1$ between Cr-Os and $J_2$ between Os-Os are both antiferromagnetic. 
We have retained next nearest neighbor Os-Os couplings because  
$5d$ orbitals are much more extended than $3d$ orbitals. 
$J_1$ tends to order the spins into a Ne\'{e}l antiferromagnet, 
while $J_2$ frustrates this ordering. Note that  
$J_2$ by itself is also frustrated since Os spins are on an FCC lattice.
Given the complexity of estimating superexchange microscopically~\cite{Fazekas},
we treat them as parameters and study the magnetic properties as a function of $J_2/J_1$. 

\smallskip

\noindent {\em {Magnetic Structure}:}
In Fig.~\ref{f3}(a), we show the magnetic ground state obtained from MC simulation of $ H_{\rm eff}$ for $J_2/J_1=0.4$. 
The magnetic structure factor has two large peaks 
at $\mathbf{q}_0=(\pi,\pi,\pi)/a$ and $\mathbf{q}_1=(\pi,\pi,0)/a$, 
where $a$ is the nearest neighbor Cr-Os separation. 
We gain insight into this ground state using a variational analysis. Fourier transforming (\ref{eq:spinH}) we get $H_{\rm eff}=\sum_q \bm{\psi}^{\dagger}(\mathbf{q})\cdot H(\mathbf{q})\cdot \bm{\psi}(\mathbf{q})$ where $H_{11}=0$, $H_{12}=H_{21}=J_1(\cos(q_xa)+\cos(q_ya)+\cos(q_za))$, $H_{22}=J_2(\cos((q_x-q_y)a)+\cos((q_y-q_z)a)+\cos((q_x-q_z)a))$ and $\bm{\psi}^{\dagger}=(\mathbf{S}^1(\mathbf{q})^*,\mathbf{S}^2(\mathbf{q})^*)$. For each $\mathbf{q}$ we diagonalize $H(\mathbf{q})$, and find
that $\mathbf{q}_0=(\pi,\pi,\pi)/a$, the usual antiferromagnetic N\'eel ordering of Cr and Os spins, minimizes the energy. While it satisfies 
$\lvert \mathbf{S}^{1}_i\rvert^2+\lvert \mathbf{S}^{2}_i\rvert^2=2S^2$, it does not satisfy the constraints
$\lvert \mathbf{S}^{1}_i\rvert^2=S^2$ and $\lvert \mathbf{S}^{2}_i\rvert^2=S^2$ for $J_2/J_1 \neq 0$.

Upon generalizing the solution to include two different $\mathbf{q}$'s
that minimize the energy and also satisfy the constraints, we find that 
$\mathbf{q}_0=(\pi,\pi,\pi)/a$ and $\mathbf{q}_1=(\pi,\pi,0)/a$ 
in agreement with the MC results.  Our prediction of the structure factor peaks 
can be checked by neutron scattering, and should prove to be an important 
test of our theory.
 
\begin{figure}[!t]
\vspace{.2cm}
\centerline{
\includegraphics[width=6cm,clip=true]{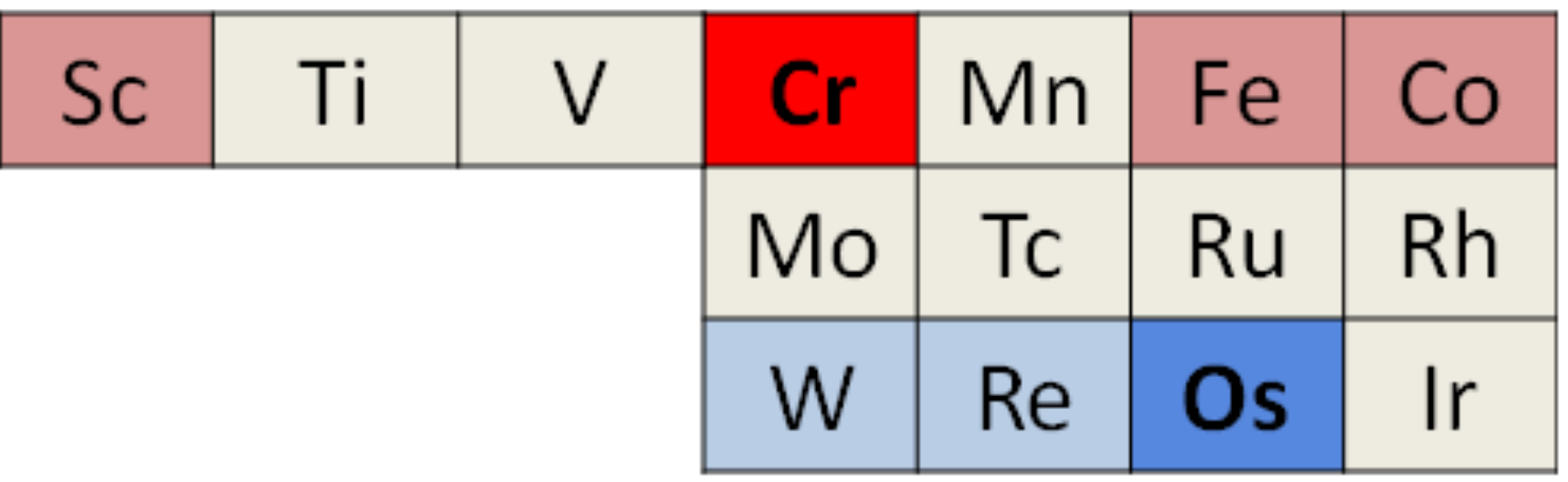}
}
\label{table1}
\end{figure}

\begin{table}

\renewcommand{\tabcolsep}{0.02cm}

\renewcommand{\arraystretch}{1.3}

\begin{tabular}{|c|c|c|c|c|c|c|c|}

\hline

~ & & & T$_c$(K) & $\theta$(K) & \multicolumn{1}{r}{} & ~ & Ref. \\ \hline

Sr$_2$ScOsO$_6$ & I & AF & 25 & -636 & Sc$^{3+}$(3d$^0$) & Os$^{5+}$(5d$^3$) & \cite{Choy_1998} \\ \hline

\bf{Sr$_2$CrOsO$_6$} & \bf{I} & \bf{Fi} & \bf{725} & \bf{-} & \bf{Cr$^{3+}$(3d$^3$)} & \bf{Os$^{5+}$(d$^3$)} & \cite{Krockenberger} \\ \hline

Sr$_2$FeOsO$_6$ & I & Fi & 65 & +37 & Fe$^{3+}$(3d$^5$) & Os$^{5+}$(5d$^3$) & \cite{Morrow} \\ \hline

Sr$_2$CoOsO$_6$ & I & Fi & 65,110 & - &Fe$^{2+}$(3d$^7$) & Os$^{6+}$(5d$^2$) & \cite{Morrow} \\

% & & & 110 & & & & \\

\hline

\multicolumn{8}{c}{ } \\

\hline

Sr$_2$CrWO$_6$ & M & Fi & 458 & - & Cr$^{3+}$(3d$^3$) & W$^{5+}$(5d$^1$) & \cite{Philipp_2003} \\ \hline

Sr$_2$CrReO$_6$ & ? & Fi & 635 & - & Cr$^{3+}$(3d$^3$) & Re$^{5+}$(5d$^2$) & \cite{Kato_2002,hauser_2012}\\

\hline

\multicolumn{8}{c}{ } \\

\hline

NaOsO$_3$ & I & AF & 410 &- &\multicolumn{1}{r}{Os$^{5+}$(5d$^3$) }& ~ &\cite{Calder_arxiv2012}\\ \hline

LaCrO$_3$& I & AF & 320 &-750 &\multicolumn{1}{r}{Cr$^{3+}$(3d$^3$) }& ~ & \cite{Koehler_1957,Sardar_2011}\\ \hline

\end{tabular}

\caption{
%Double and ternary perovskites with 
Oxides of Cr and Os, where I denotes Insulators, M metals, AF antiferromagnets, and Fi ferrimagnets.
The Weiss constant $\theta$ is not available for materials with high $T_c$. }

\end{table}

\smallskip

\noindent {\em {Magnetization $M(T)$}:}
We calculate $M(T)$ in 3D using finite temperature classical MC simulations. 
Fig. \ref{f3}(c) shows that (i) $M(T=0)$ is finite, (ii) $M(T)$ has an unusual non-monotonic function of temperature,
(iii) a large value of $T_c$ for $J_1 = 35$ meV and $J_2/J_1 = 0.4$.
Remarkably, all of these results compare very well with experimental data shown in the inset of Fig.~\ref{f3}(c) \cite{Krockenberger}. 

The non-zero value of $M(0)$ arises from the canted nature of the ground state discussed above; see Fig.~\ref{f3}(a).
There are some quantitative differences with experiment, e.g., $M(T=0)\approx 0.75 \mu_B$ and it
peaks at $\approx 0.9\mu_B$ while the simulation give $M(T=0)\approx 0.5\mu_B$, which could be due to 
any one of several effects ignored in our minimal model, like higher order effects of SOC 
or distortion from the cubic lattice.

The non-monotonicity of the net $M(T)$ arises from a competition between the magnetizations on the two sublattices:
for essentially antiparallel moments $M=M_{\rm Cr}-M_{\rm Os}$.
At low $T$, the Os moments are stuck in the canted state and $M_{\rm Os}$ does not change significantly with temperature. However 
$M_{\rm Cr}$ drops faster leading to a decrease in $M$.
This trend continues till about 150K. At intermediate temperatures, Os spins have a smaller spin stiffness compared to Cr spins because $J_2$ acts only on Os sublattice. 
Thus Os spins become more ``floppy'' and depolarize faster leading to an increase in $M$ with temperature. Close to T$_c$ both Cr and Os moments drop very rapidly and the net moment goes to zero at T$_c$. This provides a qualitative understanding of the non-monotonic behavior. The behavior shown in Fig. \ref{f3}(c) depends sensitively on the choice of  $J_2/J_1$, and $M(T)$ for different $J_2/J_1$ is 
discussed in ref.~\cite{supp}.

\noindent{\em {Discussion:}}
For weakly correlated materials, the magnetism is of the Slater/spin-density wave form where the state above
$T_c$ is a metal and the charge gap is determined by magnetic ordering,  
In a local-moment Mott insulator, the charge gap scale is set by the Coulomb interaction and is much larger
than $T_c$, set by superexchange. SCOO is in effect just on the Mott side of this
crossover between Slater and Mott magnetism. By combining a weakly interacting 5d Os with a strongly interacting 3d Cr, 
we are able to tune the interaction strength to an intermediate regime and, therefore, maximize the $T_c$. 
We already see some evidence for the emergence of such a crossover behavior: NaOsO$_3$ is a Slater
insulator with a $T_N \approx 410$ K, while LaCrO$_3$ a strong coupling Mott insulator with a $3.5$ eV charge gap and
$T_{N}\approx 320$ K. SCCO is at an intermediate coupling on the Mott side, and this generates a very high ordering temperature.

Moreover, the insulating and magnetic mechanisms that we have unravelled in SCOO
are an important step for understanding the trends, such as metal to insulator transitions,
in materials with fixed Os and varying 3d ion (Sc-Cr-Fe-Co), 
as well as fixed Cr and varying 5d ion (W-Re-Os) [see Table 1].

\noindent{\em {Conclusion:}} 
We have established a general theoretical framework for understanding Mott insulators in DP's with 
half-filled bands. It can be easily generalized to other materials like Sr$_2$MnOsO$_6$, 
Sr$_2$FeOsO$_6$~\cite{note3}. 
An important open question is the effect of doping these multi-band Mott insulators. 
Away from half-filling, SOC will become important, and the interplay of strong SOC and interactions may produce exciting new physics. 

\noindent{\em{Acknowledgments:}}  We thank Anamitra Mukherjee
for fruitful discussions. 
Our research was supported by the Center for Emergent Materials, an NSF MRSEC (Award Number DMR-0820414).


\begin{thebibliography}{}
\bibitem{note2}
   The only DFT calculation~\cite{Kubler} that estimates $T_c$, 
   ignores SOC, yields an antiferromagnetic ground state with zero net moment, 
   and an order-of-magnitude larger gap compared with other DFT calculations
   ~\cite{DDSarma,Pickett1}.
\bibitem{note3}
  Substituting Cr with Mn(Fe) will change $\Delta$ and filling. Our theory can 
  account for $\Delta$, while the extra electrons will go to the $e_g$ band. The 
  large charge transfer gap between Os $e_g$ and Mn(Fe) $e_g$ levels leads to a 
  narrow $e_g$ band which can be easily localized.
\end{thebibliography}
\end{document}